\documentclass[amsmath,aps,prb,superscriptaddress,showpacs,amssymb,secnumarabic,eqsecnum,twocolumn,floatfix]{revtex4-1}
\usepackage{graphicx}
\usepackage{pstricks}
\usepackage{hyperref,url}
\usepackage{dsfont}
\def\ba{\begin{eqnarray}}
\def\ea{\end{eqnarray}}
\def\be{\begin{equation}}
\def\ee{\end{equation}}

\graphicspath{{graficos/}}

\def\ba{\begin{eqnarray}}
\def\ea{\end{eqnarray}}
\def\be{\begin{equation}}
\def\ee{\end{equation}}

\begin{document}

\title{REENTRANT BEHAVIOUR IN LANDAU FERMI LIQUIDS WITH SPIN-SPLIT POMERANCHUK INSTABILITIES}
\author{P. Rodr\'\i guez Ponte}
\affiliation{Instituto de F\'\i sica de La Plata and Departamento de F\'isica, Universidad Nacional de La Plata, C.C. 67, 1900 La Plata, Argentina}
\author{D.C. Cabra}
\affiliation{Instituto de F\'\i sica de La Plata and Departamento de F\'isica, Universidad Nacional de La Plata, C.C. 67, 1900 La Plata, Argentina}
\author{N. Grandi}
\affiliation{Instituto de F\'\i sica de La Plata and Departamento de F\'isica, Universidad Nacional de La Plata, C.C. 67, 1900 La Plata, Argentina}
\affiliation{Abdus Salam International Centre for Theoretical Physics, Associate Scheme \\ Strada Costiera 11, 34151, Trieste, Italy}

\begin{abstract}
We study the effects of spin-antisymmetric interactions on the stability of a Landau-Fermi liquid on the square lattice, using the generalized Pomeranchuk method for
two-dimensional lattice systems. In particular, we analyze interactions that could induce instabilities of the so called spin-split type, that is when
spin-up and spin-down Fermi surfaces are displaced with respect to each other. The phase space is
studied as a function of the strength of the interaction $V$, the electron chemical potential $\mu$ and an external magnetic field $h$.
We find that such interactions produce in general an enhancement of the instability region of the Landau-Fermi liquid. More interestingly, in certain
regions of the $V$-$\mu$ phase space, we find a reentrant behaviour as a function of the magnetic field $h$, similar to that found in recent experiments,
{\it e.g.} in URu$_2$Si$_2$ and Sr$_3$Ru$_2$O$_7$.
\end{abstract}
\pacs{75.10 Jm, 75.10 Pq, 75.60 Ej}
\maketitle

\section{Introduction}
The low energy physics of weakly interacting fermions in three dimensions is mostly described by the Landau theory of the Fermi Liquid which is based on the existence of single particle fermionic excitations, or quasiparticles, with a long lifetime at very low energies\cite{Legget,Baym,Pomeranchuk}. However, moving away from three dimensions, the situation changes drastically: in one dimensional systems Landau's
quasiparticles are typically unstable, giving rise to the so called Luttinger liquid\cite{Haldane}. On the other hand, two dimensional lattice models are far more complicated to treat, since conventional perturbation theory is not applicable, and many instabilities can occur$^{5-26}$. 

The appearance of broken symmetry phases due to electron interactions within the Landau description of a Fermi liquid has been the subject of numerous investigations\cite{Quintanilla3,Jaku,Wu,Wuotro,Bonfim}. In particular, Pomeranchuk instabilities leading to nematic phases have been analized by Yamase and Kohno in Ref. \onlinecite{yamase}. Of special interest are those phases that are still described by Landau's theory, such as metallic ferromagnetism or incommensurate density waves. In ref. \onlinecite{Hirsch1} a new type of instability was discussed that lead to a phase in which the two Fermi surfaces for up and down spins get a relative shift in momentum space. This so-called spin-split phase (or $\alpha$-phase\cite{Wu}), which breaks reflection symmetry but preserves time reversal (as opposed to a ferromagnetic metallic instability) was shown to appear in the presence of interactions that arise naturally in a tight binding description of interacting electrons. The same author have analyzed the 
competition of this kind of interactions with Rashba and Dresselhaus ones\cite{Hirsch2010}. 

The subject has regained interest more recently, due to its possible relevance in the description of different phase transitions observed experimentally, such as the so-called hidden order transition in URu$_2$Si$_2$. More precisely, in ref. \onlinecite{varma-2006} the possible relevance of spin-antisymmetric interactions in the stability of the Fermi liquid and the relation with this hidden order transition has been discussed. Recent measurements, showing rotational symmetry breaking in the hidden-order phase\cite{Okazaki}, motivated the proposal of a nematic emergent phase\cite{Fujimoto}, in contrast to previous scenarios\cite{varma-2006}.

It is the purpose of this paper to analyze further this issue by applying a method that has
been recently developed in refs. \onlinecite{Lamas1}, \onlinecite{Lamas2}, \onlinecite{Lamas3}, which is specially suited
to study the Fermi liquid instabilities in two-dimensional lattice systems. This technique allows
to analyze the interplay between an external magnetic field and the underlying lattice.
In particular, we apply the extension of the above mentioned method to the zero-temperature\cite{tempdiscuss} and spinful case \cite{Lamas3} to the
class of interactions studied in \onlinecite{varma-2006}.

Our results show two important characteristic features. On the one hand
Pomeranchuk instabilities appear when the Fermi surface deformations are of the spin-split type and this leads in general to an enhancement of the instability region. On the other hand, and more importantly, with this kind of deformations a reentrant behaviour in a magnetic field is observed, similar to what is observed in {\it e.g.} URu$_2$Si$_2$ and Sr$_3$Ru$_2$O$_7$. These results should be compared with the analysis presented in \onlinecite{YK} where reentrant behaviour was observed analyzing nematic deformations of the Fermi surface. Here we obtain a similar behavior, 
without imposing nematic deformations, for a different kind (spin-split) of Fermi surface deformations.

The paper is organized as follows: in Section \ref{sec:model} we present the Hamiltonian of the system and its description as a Landau Fermi liquid, in Section \ref{sec:method} we briefly review the generalized Pomeranchuk method for lattice systems\cite{Lamas1,Lamas2,Lamas3}, and in Section \ref{sec:results} we present the main results of this paper. Finally in Section \ref{sec:conclusions} we discuss the conclusions and outlook.
\section{The model}
\label{sec:model}
We study a system of fermions in a two-dimensional square lattice with a spin-antisymmetric interaction\cite{varma-2006}. The Hamiltonian is given by
\small
\be
\hat{H}\!=\!
\sum_{{\bf k},\alpha}\varepsilon({\bf k})
\hat{c}^\dagger_{{\bf k}\alpha} \hat{c}_{{\bf k}\alpha}
+
\frac{1}{2} V\!\!\!\!\!\!
\sum_{
\substack{{\bf{k},\bf{k}',i}\\
{\alpha,\beta,\alpha',\beta'}}
}\!\!\!\!\!
\mathsf{J}_{{\bf k}{\bf k}'}
(\hat{c}^\dagger_{{\bf k}\alpha}\sigma_{\alpha\beta}^i\hat{c}_{{\bf k}\beta})
(\hat{c}^\dagger_{{\bf k}'\!\alpha'}\sigma_{\alpha'\!\beta'}^i \,\!\hat{c}_{{\bf k}'\!\beta'}) \,,
\label{H}
\ee
\normalsize
where  $\sigma^i$ are the Pauli matrices and $\alpha=\pm1/2$ denotes the spin $z$-projection.
We consider in this paper the dispersion relation $\varepsilon({\bf k})$ which
corresponds to nearest neighbor hopping
\be
\varepsilon({\bf k})=-2t(\cos{k_x}+\cos{k_y})
\label{eq:disprel}\,,
\ee
the analysis of more general cases being straightforward.

For the interaction, following\cite{varma-2006} we choose $\mathsf{J}_{{\bf k}{\bf k}'}$  in the antisymmetric spin channels
\be
 \mathsf{J}_{{\bf k}{\bf k}'}=\sum_{l} \mathsf{J}_l P_l (\cos{\theta _{{\bf k}{\bf k}'}})\,.
\ee
where $\theta _{{\bf k}{\bf k}'}$ is the angle between ${\bf k}$ and ${\bf k}'$.
For the sake of simplicity we consider in this paper a single $l$-channel, which we fix to $l=1$,
but our analysis can be easily generalized to higher $l$-channels.

We take into account the presence of an external magnetic field $h$ by adding to (\ref{H}) the Zeeman interaction
\be
\hat{H}_{mag}=2h \sum_{{\bf k},\alpha} \alpha\, \hat{c}^\dagger_{{\bf k}\alpha} \hat{c}_{{\bf k}\alpha}\,.
\ee

To carry the analysis further, we assume that the above system has a Fermi liquid phase that can be described using Landau's theory. Then we can write the change in the Landau free energy $\Omega$, associated with a change $\delta n_{{\bf k}\alpha}$ in the quasi-particles distribution function $n_{{\bf k}\alpha}$, as
\small
\ba
\label{eq:deltaE}
 \delta \Omega\! &=&\!
\sum_{\alpha} \!\!\!\int\!\! d^2\!\!k\, (\varepsilon({\bf k})\!\!-\!\!\mu_{\!\alpha}\!) \delta n_{{\bf k}\alpha}\!\!+\!\! 
  \frac{1}{2}\!  \sum_{\alpha,\alpha\!'\!}  \!\!\int\!\!\!d^2\!k d^2\!k'\! f_{\alpha\alpha\!'}\!({\bf k},\!{\bf  k}\!') \delta n_{{\bf k}\alpha} \delta
  n_{{\bf k\!' }\!\alpha\!'}
\nonumber\\
\ea
\normalsize
where we have defined a spin dependent chemical potential by reabsorbing the external magnetic field
\be
\mu_\alpha=  \mu+2\alpha h
\ee
Within the mean field approximation\cite{Quintanilla1}, the dispersion relation in (\ref{eq:deltaE})
can be assumed to take the same form as in (\ref{eq:disprel}) with renormalized coefficients.
On the other hand, the ``interaction function'' $f_{\alpha\alpha'}({\bf k},{\bf k'})$ takes the form
\be
f_{\alpha\alpha'}({\bf k},{\bf k}')=
 2 V \mathsf{J}_{{\bf k}{\bf k}'}(2\delta_{\alpha\alpha'} -1)
\label{eq:intfunct}
\ee
The hypothesis of existence and stability of a Fermi liquid phase can now be tested: an instability is diagnosed whenever an excitation $\delta n_{{\bf
k}\alpha}$ leads to a negative Landau free energy (\ref{eq:deltaE}). The way of parameterizing the space of excitations and exploring it in the search
for possible instabilities is the core of the generalized Pomeranchuk method, and will be sketched in the next Section.
\section{The method}
\label{sec:method}
In this section we give a brief description of the generalization of Pomeranchuk's method to lattice systems, that was originally proposed in
ref. \onlinecite{Lamas1} and further developed in \onlinecite{Lamas2} and \onlinecite{Lamas3}.
Here we concentrate in the spin $1/2$ case
at zero temperature, with a dispersion relation (\ref{eq:disprel}) and interaction function (\ref{eq:intfunct}) and we refer the reader to ref. \onlinecite{Lamas3} for details.

The ground state of a spin $1/2$ system at zero temperature is described by its Fermi surfaces. With the help of the function $g_\alpha(\textbf{k})
\equiv \mu_\alpha - \varepsilon(\textbf{k})$, the Fermi surface for each spin component can be defined as the loci in momentum space of the points at
which $g_\alpha(\textbf{k}) $ vanishes
\be\label{eq:cambiovar}
g_\alpha(\textbf{k}) =0\,.
\ee
The occupation number $n_{{\bf k}\alpha}$ vanishes outside the corresponding Fermi surface and takes the value $1$ inside.

At zero temperature, an excitation of the system $\delta n_{{\bf k}\alpha}$ takes the values $0,\pm 1$ at different ${\bf k}$ points, and
in consequence it can be characterized by a deformed Fermi surface. The latter can be represented by a perturbed version of equation
(\ref{eq:cambiovar})
\be\label{eq:decompo}
g_\alpha(\textbf{k})  \!+\!\delta g_\alpha(\textbf{k}) =0
\ee
~

\noindent In order to find a useful parameterization for $\delta g_\alpha(\textbf{k})$, we change variables from $(k_x,k_y)$ to $(g_\alpha,s_\alpha)$,
where $s_\alpha=s_\alpha({\bf k})$ is defined as orthogonal to $g_\alpha({\bf k})$, {\em i.e.} tangential to the corresponding Fermi surface, and
performing a complete turn around it while running on the interval $[-\pi,\pi)$. The associated Jacobian reads
\ba
J_\alpha(s_\alpha)&=&\left.\frac{\partial(k_x,k_y)}{\partial (g_\alpha,s_\alpha)}\right|_{g_\alpha=0}=
\\
&=&\left({2t\sqrt{ 1-\left(1-\frac{({\mu+2 \alpha h})^2}{16t^2}\right)\cos^2({2s})}}\,\right)^{-1},
\nonumber\ea
Now if we choose
\be
\delta g_\alpha({\bf k}) =
\! \frac1{\sum_\gamma J_\gamma^2}
\!\left(\!
\psi^1 {J_\alpha} \!+\!  \psi^2  \sum_\beta {\epsilon_{\alpha \beta}J_\beta }
\!\right),
\ee
with $\epsilon_{\alpha \beta}$ the Levy-Civita antisymmetric symbol $\epsilon_{11}=\epsilon_{22}=0$, $\epsilon_{12}=-\epsilon_{21}=1$, implying that the factor that multiplies $\psi^2$ is perpendicular to $J_\alpha$ in $\alpha$ space. We have the deformations of the unperturbed Fermi surface corresponding to the $\alpha$ component ($g_\alpha=0$) parameterized in terms of two arbitrary functions $\psi^1(s_\alpha)$ and $\psi^2(s_\alpha)$ of a single variable $s_\alpha$.
The generality of this decomposition can be understood by replacing it into the expressions for the variations of the magnetization and the total number
of particles.
When $\psi^1(s)$ has no constant term it parameterizes a simultaneous deformation of both Fermi surfaces that do not change the magnetization of the
system nor the total number of particles. Stoner instabilities, due to fluctuations in the number of particles, can be also included by allowing a
constant term in $\psi^1(s)$. On the other hand $\psi^2(s)$ takes into account deformations that include spin flips, changing the total magnetization.

Inserting the above decomposition into the variation of the Landau free energy (\ref{eq:deltaE}), after some straightforward algebra\cite{Lamas3}, we get
\be
\delta \Omega=
\int\!\! ds \!\int\!\! ds' \!\sum_{a,b=1,2} \!\!\psi^a\!(s)  K^{ab} (s,s')   \psi^b\!(s')
\label{eq:deltaOmega}
\ee
As we have chosen both variables $s_\alpha$ as running in the same interval, we are allowed to
omit the index $\alpha$ in the dumb integration variables. Here $ K^{ab} (s,s')$ is a $2\times 2$ matrix kernel given by
\begin{widetext}
\begin{footnotesize}
\begin{eqnarray}
K^{11} (s,s') &=& \frac{1}{2\sum_\alpha\! J_\alpha^2(s) \sum_\alpha \!J_\alpha^2(s')} \left(
\sum_\alpha J_\alpha^3 \delta(s-s') +\sum_{\alpha,\beta}
J_\alpha^2(s) f_{\alpha\beta}(s,s')J_\beta^2(s')
\right) \nonumber \\
K^{22} (s,s') &=& \frac{1}{2 \sum_\alpha \!J_\alpha^2(s)  \sum_\alpha \!J_\alpha^2(s') } \left(
\sum_{\alpha, \beta,\gamma} \epsilon^{\alpha \beta} \epsilon^{\alpha \gamma} J_\alpha J_\beta J_\gamma \delta(s-s') \right.
\left. + \sum_{\alpha, \beta,\gamma,\delta} \epsilon^{\alpha\beta} \epsilon^{\gamma\delta}
J_\alpha(s) J_\beta(s) J_\gamma(s') J_\delta(s') f_{\alpha\delta}(s,s')
\right) \nonumber \\
K^{12} (s,s') &=& \frac{1}{2 \sum_\alpha \!J_\alpha^2(s)  \sum_\alpha \!J_\alpha^2(s') } \left(
\sum_{\alpha, \beta}  \epsilon^{\alpha \beta}  J_\alpha^2 J_\beta \delta(s-s') + \sum_{\alpha, \beta,\delta}
\epsilon^{\alpha\beta}
J_\delta(s)^{2} J_\alpha(s') J_\beta(s') f_{\delta\alpha}(s,s')
\right) \nonumber \\
K^{21} (s,s') &=& \frac{1}{2 \sum_\alpha \!J_\alpha^2(s)  \sum_\alpha \!J_\alpha^2(s')}  \left(
\sum_{\alpha, \beta} \epsilon^{\alpha \beta} J_\alpha^2 J_\beta \delta(s-s') +
\sum_{\alpha,\beta,\gamma} \epsilon^{\alpha \beta} J_\alpha(s) J_\beta(s) J_\gamma(s')^{2}  f_{\alpha\gamma}(s,s')
\right)
\end{eqnarray}
\end{footnotesize}
\end{widetext}
To find the instabilities of the Fermi liquid, we consider its energy (\ref{eq:deltaOmega}) as a quadratic form $\langle\psi\vert\psi\rangle$, acting
on the $2$-vector function  $\psi^a(s)$ that parameterize the deformations, and then search for a choice of $\psi^a(s)$ that turns it into negative
values. A simple way to do that is to apply the Gram-Schmidt orthogonalization procedure to an arbitrary starting basis $\{\psi^a_n(s)\}_{n\in
\mathds{N}}$, to obtain a new basis of orthogonal functions $\{\xi^a_n(s)\}_{n\in \mathds{N}}$ satisfying $\langle\xi_n\vert\xi_m\rangle\propto \delta_{nm}$. If at a given step of the procedure a $2$-vector
function $\xi^a_i(s)$ is found whose pseudonorm is negative, we say that we have found an instability in the $i$-th channel.
In our case, we choose the initial basis as
\be
\psi^a_n(s)=
\left\{
\begin{array}{ll}
\left( \begin{array}{c}\cos\!\left(\frac{ns}2\right)\\0\end{array} \right) & \mbox{if $n$ even and $a=1$} \\ & \\
\left( \begin{array}{c}\sin\!\left(\!\frac{(n\!-\!1)s}2\!\right) \\0\end{array} \right) & \mbox{if $n$ odd and $a=1$} \\ & \\
\left( \begin{array}{c}0\\ \cos\!\left(\frac{ns}2\right) \end{array} \right) & \mbox{if $n$ even and $a=2$}  \\ & \\
\left( \begin{array}{c}0\\ \sin\!\left(\frac{(n\!-\!1)s}2\right) \end{array} \right) & \mbox{if $n$ odd and $a=2$} \\ & \\
\end{array}
\right.
\ee
\section{Results}
\label{sec:results}
By exploring the set of possible Fermi surface deformations as described in the previous section we obtained the unstable regions in the phase space.

\begin{figure}[htp]
\begin{center}
  \includegraphics[width=0.15\textwidth]{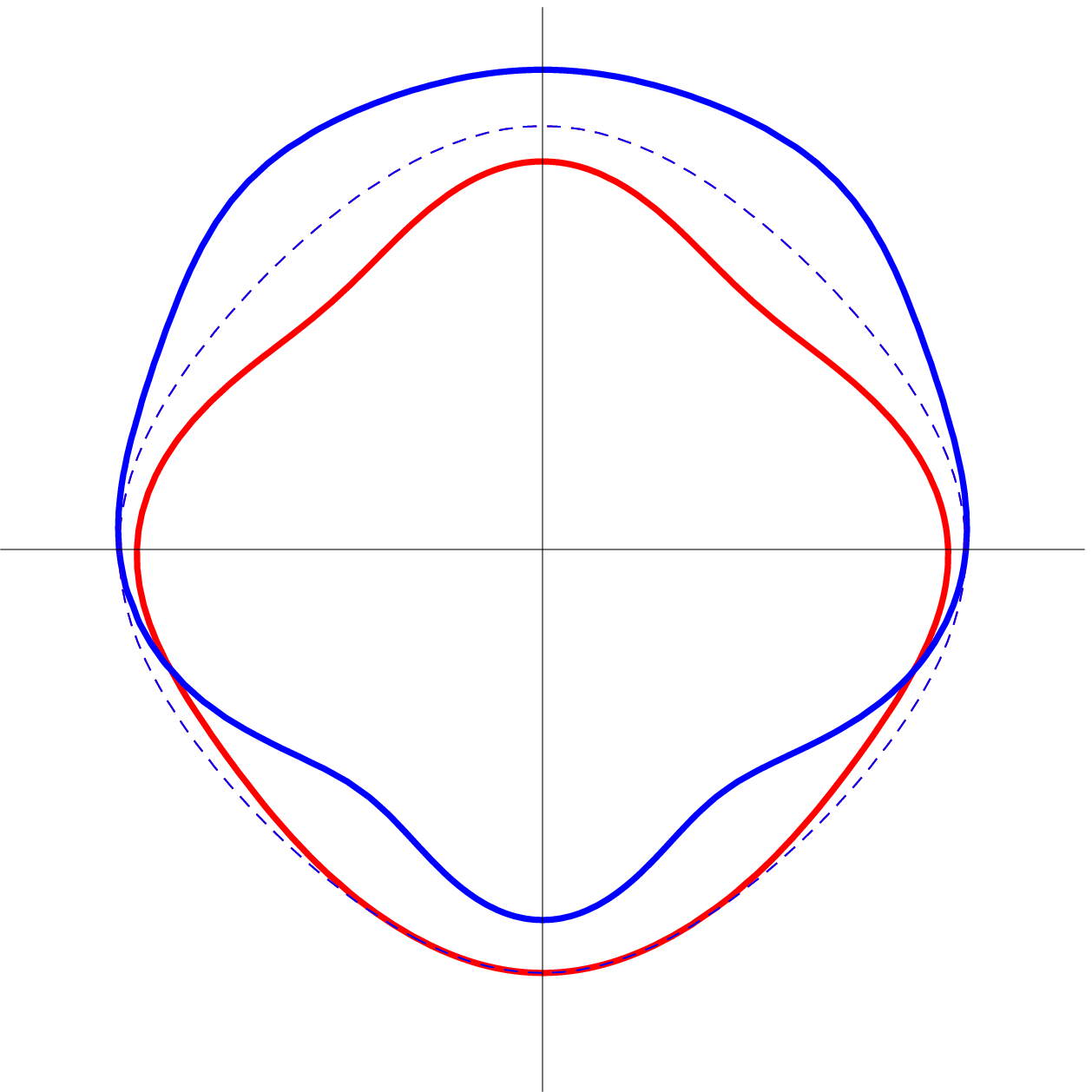}
	\begin{picture}(0,0)(100,100)
	  \put(50,175){\tiny{$k_y$}}
	  \put(92,133){\tiny{$k_x$}}
	\end{picture}
  \includegraphics[width=0.15\textwidth]{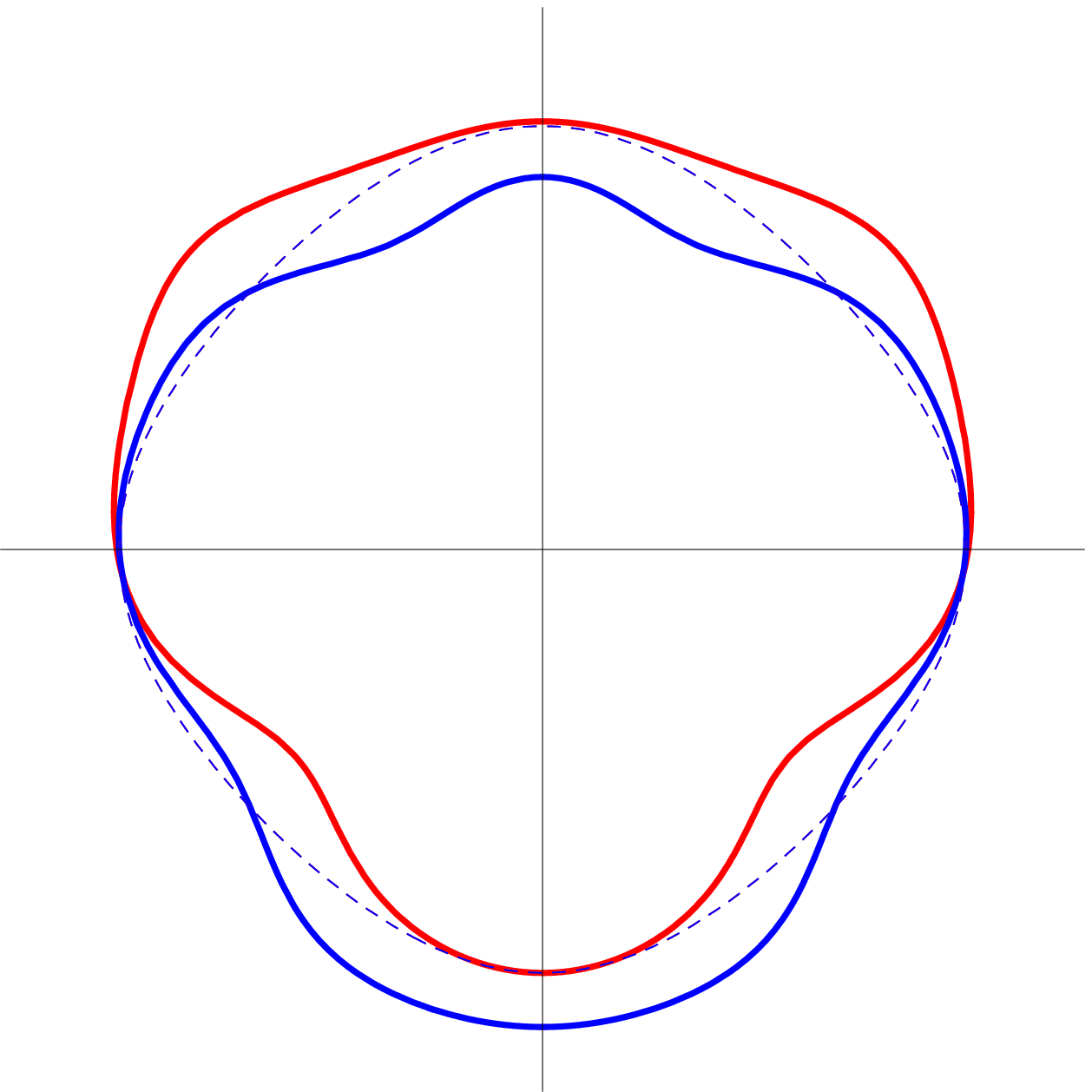}
	\begin{picture}(0,0)(100,100)
	  \put(50,175){\tiny{$k_y$}}
	  \put(92,133){\tiny{$k_x$}}
	\end{picture}
  \includegraphics[width=0.15\textwidth]{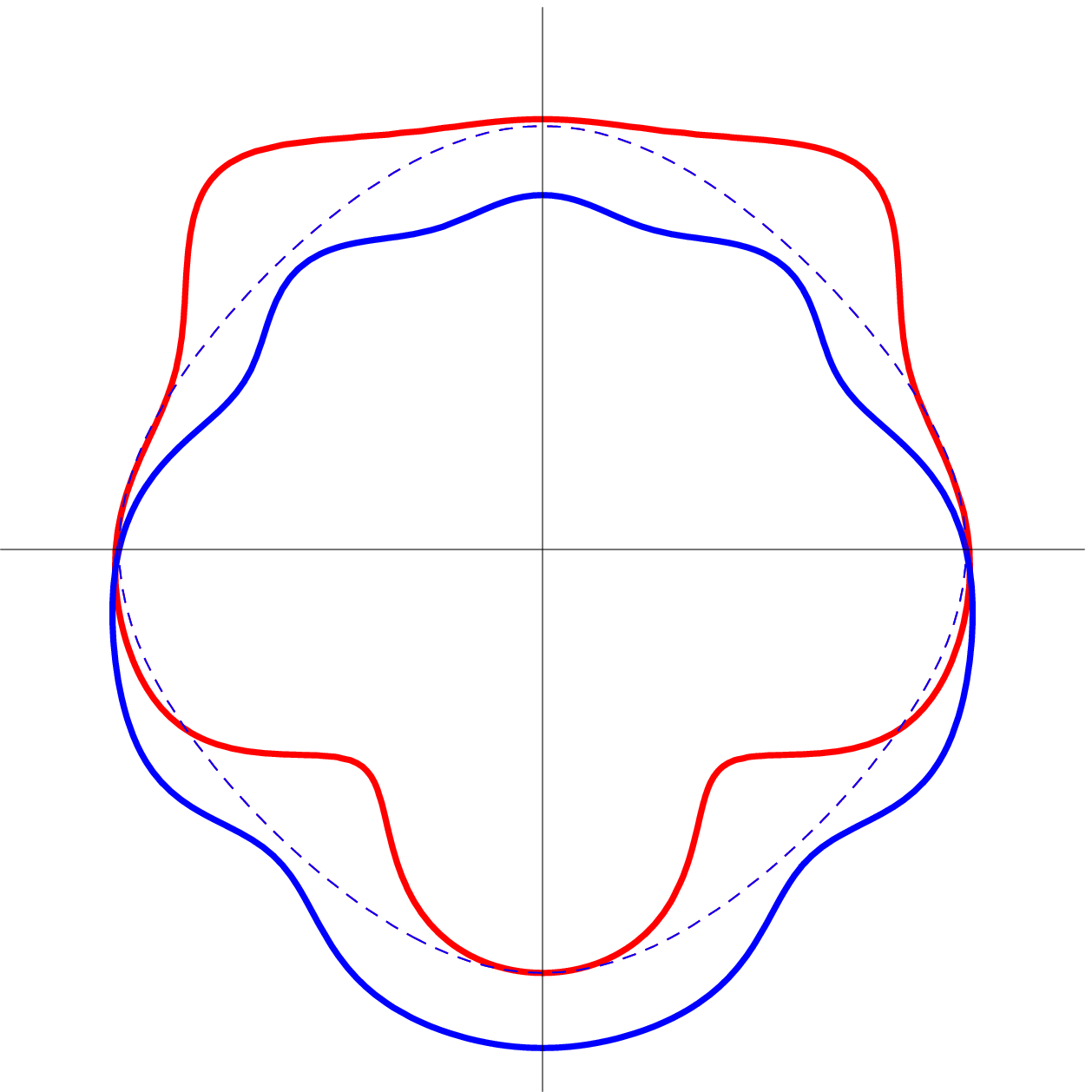}
	\begin{picture}(0,0)(100,100)
	  \put(50,175){\tiny{$k_y$}}
	  \put(92,133){\tiny{$k_x$}}
	\end{picture}
\\
\vspace{7mm}
  \includegraphics[width=0.15\textwidth]{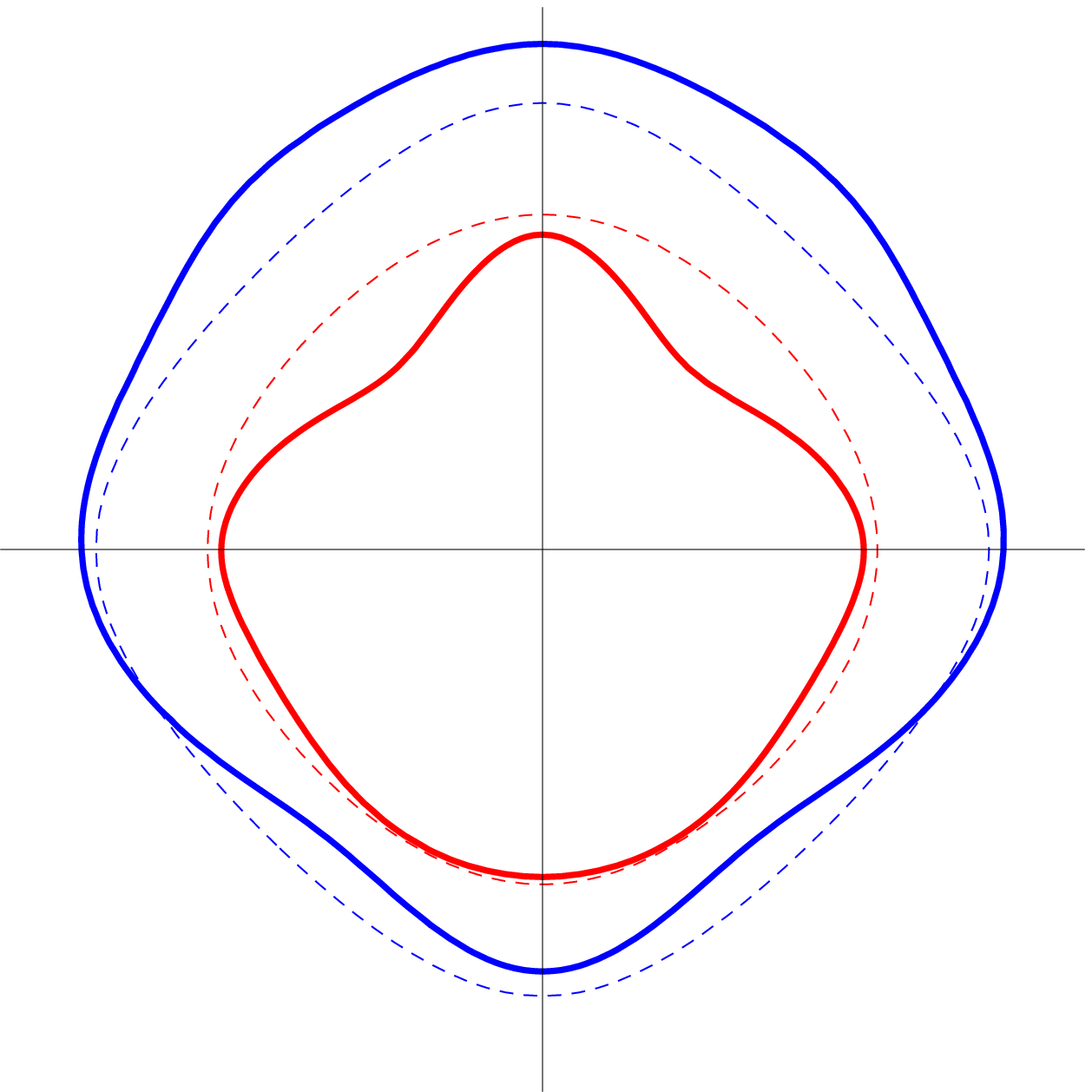}
	\begin{picture}(0,0)(100,100)
	  \put(50,175){\tiny{$k_y$}}
	  \put(92,133){\tiny{$k_x$}}
	\end{picture}
  \includegraphics[width=0.15\textwidth]{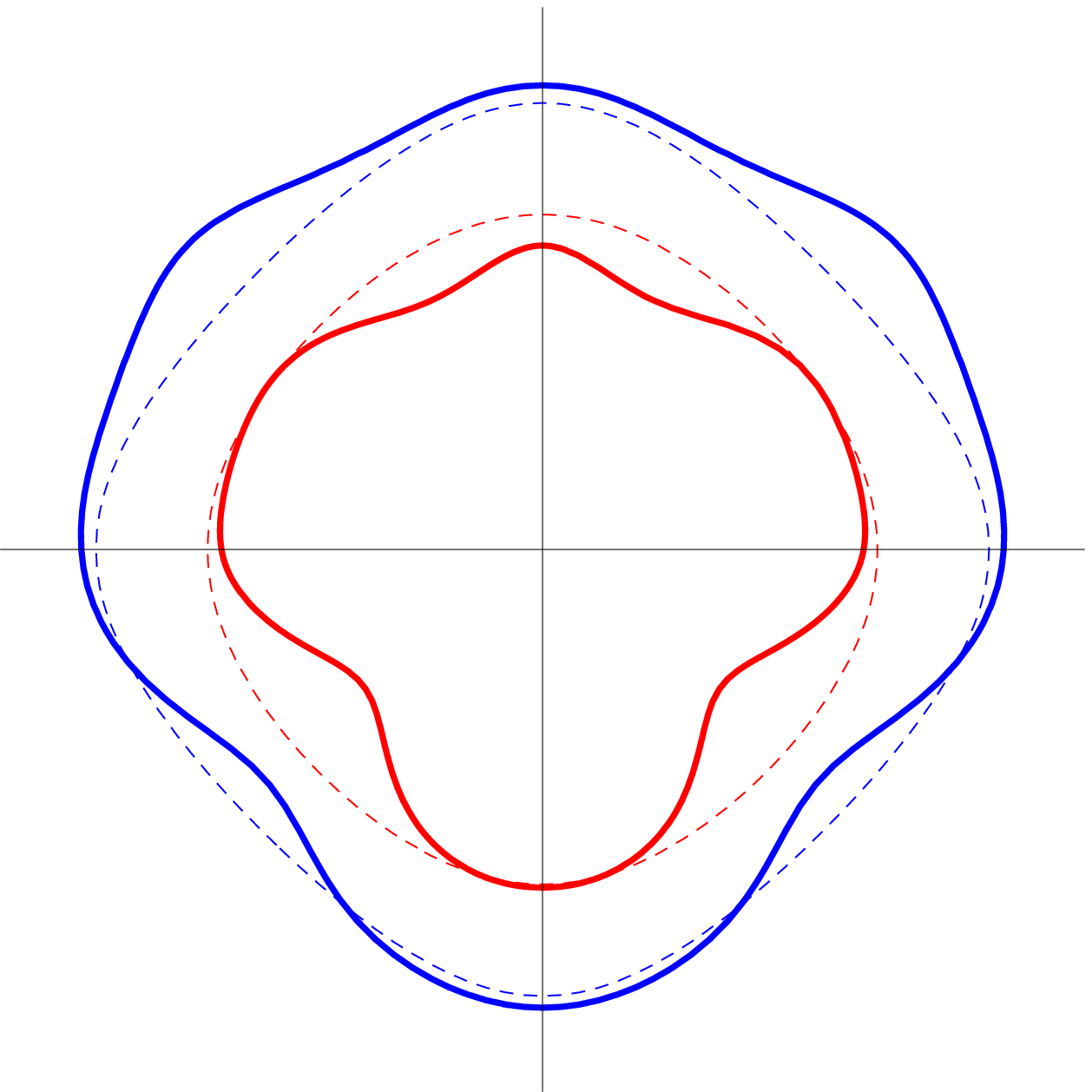}
	\begin{picture}(0,0)(100,100)
	  \put(50,175){\tiny{$k_y$}}
	  \put(92,133){\tiny{$k_x$}}
	\end{picture}
  \includegraphics[width=0.15\textwidth]{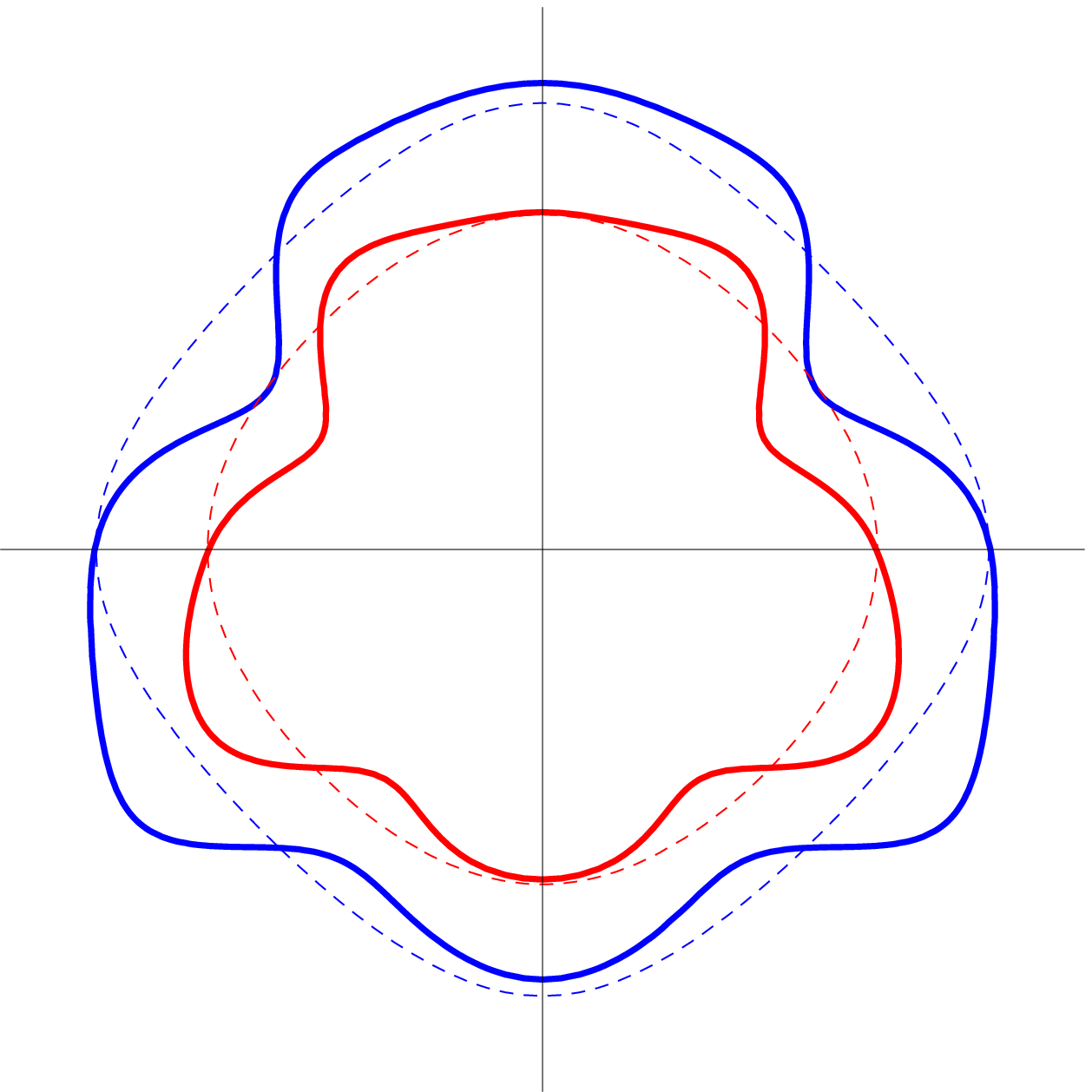}
	\begin{picture}(0,0)(100,100)
	  \put(50,175){\tiny{$k_y$}}
	  \put(92,133){\tiny{$k_x$}}
	\end{picture}
\end{center}
\caption{(Color online) Fermi surface deformations leading to instabilities of the Fermi liquid. In blue (red) we plotted the spin-up (spin-down) perturbed Fermi surfaces, the dashed lines corresponding to the unperturbed ones. The first modes for $\mu/t=-2$, $V/t=-2$ are shown, the upper row corresponding to $h/t=0$, the lower one to $h/t=0.5$. Note the relative displacement of the spin up and spin down Fermi surfaces in the first row, which indicates that the instability will probably lead to a spin-split state.}
 \label{deformaciones}
\end{figure}

In Fig.\ \ref{deformaciones} we show deformed Fermi surfaces for
both spin projections, corresponding to the first modes that lead to instabilities. As our method works with infinitesimal deformations, the surfaces
shown in the figure are schematic, since the deformation
amplitude has been augmented for the sake of illustration.

\begin{figure}
\includegraphics[width=0.4\textwidth]{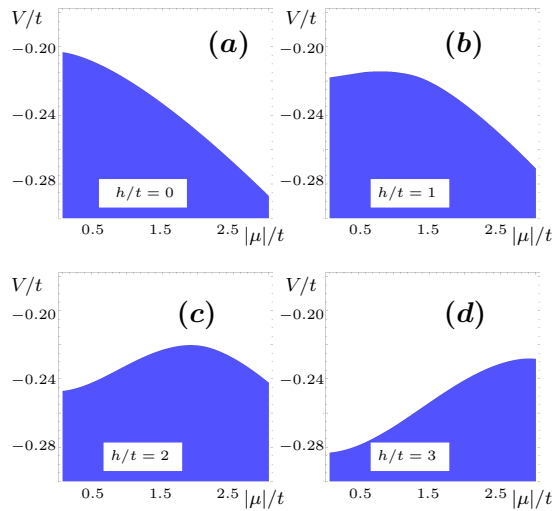}
\caption{(Color online) Unstable regions (shaded) for external magnetic field going from $ h/t=0$ to $ h/t=3$, in figs. (a) to (d).}
\label{inest}
\end{figure}

We explored an ample range of parameters $\mu/t$, $V/t$ and $h/t$ and obtained the instability regions shown in Fig.\ \ref{inest}(a)to(d).
One can observe that the instability becomes increasingly significant for $\mu=\pm h$, where one of the Fermi surfaces gets close to the van Hove singularity.
\begin{figure}
\includegraphics[width=0.4\textwidth]{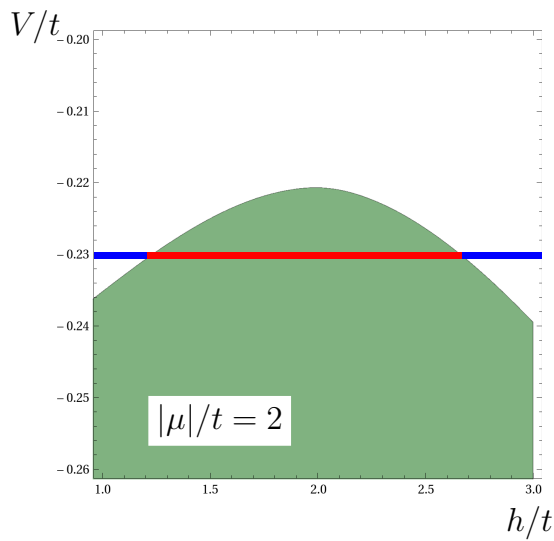}
\caption{(Color online) Instability region as a function of applied magnetic field; by increasing it, is possible to leave (red line) and re-enter (blue line) to the stable Fermi liquid phase.}
\label{inest2}
\end{figure}
A remarkable feature of the present phase diagrams is that, for some values of $\mu$ and interaction strength $V$, the Fermi liquid
description breaks down only for intermediate values of $h$. For instance, as shown in Fig.\ \ref{inest2}, if the system is filled up to $\mu/t=2$ and the interaction strength is $V/t=-0.23$, then we have a stable Fermi liquid up to $h/t\approx 1.4$ and from $h/t\approx 2.6$ but such Fermi liquid description does not hold
when $h/t$ takes values in between those limits. This means that, under such conditions, it is possible for the system to leave and re-enter the Fermi liquid phase, as the external field is increased.
\vspace{2cm}

\section{Conclusions}
\label{sec:conclusions}
In this paper we applied the method developed in \cite{Lamas1,Lamas2,Lamas3} to a lattice model with spin degrees of freedom and in the presence of interactions that break the spin symmetry\cite{varma-2006}. The phase space was studied as a function of the interaction strength $V$, the electron chemical potential $\mu$ and an external magnetic field $h$.
Our results show that Pomeranchuk instabilities appear when the Fermi surface
deformations are of the spin-split type  (that is when spin-up and spin-down Fermi surfaces are
displaced with respect each other in momentum space), confirming the studies in \onlinecite{Hirsch1} and \onlinecite{varma-2006}. This leads in general to an enhancement of the instability region.
On the other hand, and more importantly, we observe a reentrant behaviour in a magnetic field, similar to
what is observed in {\it e.g.} URu$_2$Si$_2$ and Sr$_3$Ru$_2$O$_7$ (see {\it e.g.} \onlinecite{Grigera2}).
\section*{Acknowledgments: }
We thank Carlos Lamas, Carlos Na\'on and Gerardo Rossini for helpful discussions. This
work was partially supported by the ESF grant INSTANS, PICT ANPCYT (Grants No 20350 and 00849), and PIP CONICET (Grants No 5037, 1691 and 0396).


\begin{thebibliography}{0}
\bibitem{tempdiscuss} When considering finite temperature effects, the Fermi Liquid parameters should be modified by nonanalytic leading corrections as discussed in refs. \onlinecite{Chitov} and \onlinecite{Chubukov}. In our previous work\cite{Lamas3} temperature effects were included as a low $T$ expansion and consequently only standard corrections were obtained. However this is not a problem inherent to the method we developed, but only to the low T approximation, which can be avoided. This is the subject of a future paper that is now in preparation.
\end{thebibliography}
\end{document}